\documentclass[conference]{IEEEtran}
\IEEEoverridecommandlockouts
\usepackage{cite}
\usepackage{amsmath,amssymb,amsfonts}
\usepackage{algorithm}
\usepackage{algorithmic}
\usepackage{graphicx}
\usepackage{textcomp}
\usepackage{xcolor}
\usepackage{bm,booktabs,multirow}
\usepackage{hyperref}

\def\BibTeX{{\rm B\kern-.05em{\sc i\kern-.025em b}\kern-.08em
    T\kern-.1667em\lower.7ex\hbox{E}\kern-.125emX}}
\begin{document}

\title{\textbf{EDSep: An Effective Diffusion-Based Method for Speech Source Separation}\\

\thanks{This work was supported by the National Nature Science Foundation of China under Grant 62176106, the Special Scientific Research Project of the School of Emergency Management of Jiangsu University under Grant KY-A-01, the Project of Faculty of Agricultural Engineering of Jiangsu University under Grant NGXB20240101, the Key Project of National Nature Science Foundation of China under Grant U1836220, the Jiangsu Key Research and Development Plan Industry Foresight and Key Core Technology under Grant BE2020036.

$^{\star}$ Corresponding Author.}
}

\author{
Jinwei Dong$^{1}$ \qquad 
Xinsheng Wang$^{3}$ \qquad  
Qirong Mao$^{1,2~\star}$ \\
\IEEEauthorblockA{
$^{1}$ \textit{School of Computer Science and Communication Engineering, Jiangsu University} \\
$^2$ \textit{Jiangsu Engineering Research Center of Big Data Ubiquitous Perception and Intelligent Agriculture Applications} \\
$^2$\textit{Provincial Key Laboratory of Computational Intelligence and New Technologies in Low-Altitude Digital Agriculture} \\
Zhenjiang, China \\
$^3$\textit{Audio, Speech and Language Processing Group, School of Computer Science,
Northwestern Polytechnical University} \\
Xi’an, China \\
2212208010@stmail.ujs.edu.cn, w.xinshawn@gmail.com, mao\_qr@ujs.edu.cn}
}
\maketitle

\begin{abstract}
Generative models have attracted considerable attention for speech separation tasks, and among these, diffusion-based methods are being explored. Despite the notable success of diffusion techniques in generation tasks, their adaptation to speech separation has encountered challenges, notably slow convergence and suboptimal separation outcomes. To address these issues and enhance the efficacy of diffusion-based speech separation, we introduce EDSep, a novel single-channel method grounded in score matching via stochastic differential equation (SDE). This method enhances generative modeling for speech source separation by optimizing training and sampling efficiency. Specifically, a novel denoiser function is proposed to approximate data distributions, which obtains ideal denoiser outputs. Additionally, a stochastic sampler is carefully designed to resolve the reverse SDE during the sampling process, gradually separating speech from mixtures. Extensive experiments on databases such as WSJ0-2mix, LRS2-2mix, and VoxCeleb2-2mix demonstrate our proposed method's superior performance over existing diffusion and discriminative models, validating its efficacy.

\end{abstract}


\section{Introduction}
Speech separation is a technology that extracts desired signals from mixed audio. This task involves two primary data types: single-channel speech \cite{luo2018tasnet} and multi-channel speech \cite{luo2020end}. Single-channel speech is characterized by the absence of multiple data sources, lacking the spatial information in multi-channel speech, which captures sound from various directions. This absence of spatial cues significantly complicates the task of isolating clear speech from single-channel recordings, thereby garnering considerable attention in the research community. In this work, we also focus on the separation in single-channel speech, aiming to advance the application of diffusion processes in this challenging context.

Discriminative methods have been pivotal in the field of speech separation, particularly after the introduction of deep clustering~\cite{hershey2016deep} and permutation invariant training (PIT)~\cite{kolbaek2017multitalker}, which effectively resolved the label permutation issue. Initially, research predominantly concentrated on estimating magnitudes, with or without phase consideration~\cite{wang2018end, wang2019deep}. Later developments shifted towards performing separation directly in the complex time-frequency (T-F) domain using complex ratio masking~\cite{liu2019divide}, or in the time domain through TasNets~\cite{luo2018tasnet, luo2019conv}. Among these, Conv-TasNet~\cite{luo2019conv}, employing the encoder-separator-decoder framework, has emerged as a widely adopted method for time-domain speech separation~\cite{zeghidour2021wavesplit, rixen2022qdpn, qian2022efficient,rixen2022sfsrnet}. To manage extremely, long encoded input sequences, DPRNN~\cite{luo2020dual} introduces an efficient dual-path framework that processes data by dividing it into smaller chunks. The performance is further enhanced by VSUNOS~\cite{nachmani2020voice}, which integrates gated RNN modules. Recently, the Transformer architecture, which is based on self-attention mechanisms~\cite{vaswani2017attention}, has been successfully integrated into the dual-path speech separation pipeline~\cite{subakan2021attention, chen2020dual}. MossFormer~\cite{zhao2023mossformer} is built on a gated single-head Transformer that incorporates convolution-augmented joint self-attentions.

Generative methods have recently garnered significant attention in the domain of speech separation. Unlike discriminative approaches, generative models focus on training to approximate complex data distributions and use mixed speech as input to produce clean speech outputs. Li et al.~\cite{li2021generative} introduced a method employing a generative adversarial network (GAN) tailored for speech separation. Their network is designed with dual objectives: suppressing reverberation and enhancing target speech. Similarly, Do et al.~\cite{do2020speech} developed a speech separation technique based on a variational autoencoder (VAE), which efficiently extracts and enhances the speech signal from mixed speech inputs.

Over the past several years, diffusion-based models \cite{bond2021deep} have achieved rapid and substantial advancements in the domains of image generation \cite{ramesh2022hierarchical, rombach2022high} and audio generation \cite{kong2020diffwave, xue2022learn2sing}. More recently, it has been applied to the field of speech separation. A notable example is DiffSep \cite{scheibler2023diffusion}, which employs a carefully formulated stochastic differential equation (SDE), and trains the SDE model to convert separated signals back to their mixtures, allowing the reverse process to isolate individual sources from the mixed signal effectively.


Despite recent advancements, diffusion-based methods in speech separation tasks suffer from long training periods and suboptimal quality of the clean speech produced, particularly in real-world scenarios. There remains a performance gap compared to SOTA discriminative methods. To bridge this gap, we propose EDSep, an SDE-based method that employs a variance exploding SDE tailored to the characteristics of distinct signals, which typically converge towards their collective mixture. To enhance both the efficiency and quality of speech separation, we introduce an improved training and sampling process. Moreover, drawing inspiration from DiffSep~\cite{scheibler2023diffusion}, we refine the training strategy to mitigate the initial ambiguity in source assignment during the sampling process, thereby enabling more precise identification and allocation of sources.


\section{Background}

\subsection{SDE for Diffusion-Based Modelling}
Diffusion-based modeling through SDE is an effective method for simulating complex data distributions \cite{song2020score}. It contains a forward and a reverse process. The forward process initiates with samples from the target distribution and progressively converges to a Gaussian distribution, which can be mathematically described by
\begin{equation}
    d\bm{x}=f(\bm{x},t)dt+g(t)d\bm{w} \textit{,}
\label{forward}
\end{equation}
where $\bm{x}$ is a function with vector values that depends on time $t$, and $d\bm{x}$ represents its derivative with respect to time $t$. Here, $dt$ denotes an infinitesimal time step. The functions $f$ and $g$ are known as the drift and diffusion coefficients, respectively, which govern the dynamics of $\bm{x}$. The term $d\bm{w}$ represents a standard Wiener process \cite{sarkka2019applied}. The primary goal of the forward process is to systematically introduce incremental noise into the training data by manipulating different configurations of $f$ and $g$, ensuring that the prior distribution $p_{t}(\bm{x}) $ closely approximates a Gaussian distribution.

Given certain moderate prerequisites, the SDE in Eq.~\ref{forward} has an associated reverse-time SDE:
\begin{equation}
    {d}\bm{x}=[f(\bm{x},t)-g(t)^{2}\nabla _{\bm{x}}\log{p_{t}}{(\bm{x})}]{d}t+g(t){d}{\bar{\bm{w}}} \textit{,} 
\end{equation}
which operates from $t = T$ down to $t = 0$~\cite{anderson1982reverse}. The term $d{\bar{\bm{w}}}$ is a reverse Brownian process, and $p_{t}(\bm{x})$ is the marginal distribution of $\bm{x}_{t}$.

In the forward process, starting from a known initial state $x_0$, it is typically feasible to obtain an explicit formula for $p_{t}(\bm{x})$. However, the reverse process poses a challenge because $p_t(\bm{x})$ is typically unknown, making direct solutions difficult. The key to solving this problem lies in training a neural network to approximate $\nabla _{\bm{x}}\log{p_{t}}{(\bm{x})}$.

\subsection{Denoising Score Matching}
The score function, a derivative of the log probability density with respect to the data, offers a notable advantage in that it does not depend on the often intractable normalization constant of the probability density function $p(\bm{x}; \sigma)$. This characteristic makes the score function easier to evaluate and apply in practical settings. 
For the sake of completeness, Karras et al. \cite{karras2022elucidating} derive the connection between score matching and denoising for a finite dataset. They design a denoiser function as:
\begin{equation}
    D_{\theta}(\bm{x};\sigma)=c_{skip}(\sigma){\bm{x}}+c_{out}(\sigma)F_{\theta} \textit{,}
\label{denoiser}
\end{equation}
\begin{equation}
    F_{\theta}=F_{\theta}(c_{in}(\sigma){\bm{x}};c_{noise}(\sigma)) \textit{,} 
\end{equation}
where $F_{\theta}$ is the neural network to be trained, $c_{skip}({\sigma})$ modulates the skip connection, $c_{in}({\sigma})$ and $c_{out}({\sigma})$ scale the input and output magnitudes, respectively, and $c_{noise}({\sigma})$ maps noise level $\sigma$ into a conditioning input for $F_{\theta}$. The primary goal of the denoiser function  $D_{\theta}(\bm{x};\sigma)$ is to minimize the expected $L_{2}$ denoising error for samples taken from independently for each value of $\sigma$, i.e.,
\begin{equation}
    \mathbb{E}_{\bm{y} \sim p_{\textit{data}}}\mathbb{E}_{\bm{n} \sim {\mathcal{N}}(0,{\sigma}^2\mathrm{I})}{[\lambda(\sigma)\Vert D(\bm{y}+\bm{n};\sigma)-\bm{y}\Vert^2_2]} \textit{,} 
\end{equation}
where $\bm{y}$ is training data, $\lambda(\sigma)$ is the corresponding weight and $\bm{n}$ is the noise component. This approach strategically separates the noise from the actual signal in the data, leveraging the score function's capabilities to refine the signal reconstruction process effectively.

\subsection{Related Work in Diffusion-Based Speech Separation}
\begin{figure}[t]
        \centering
        \includegraphics[width=1\linewidth]{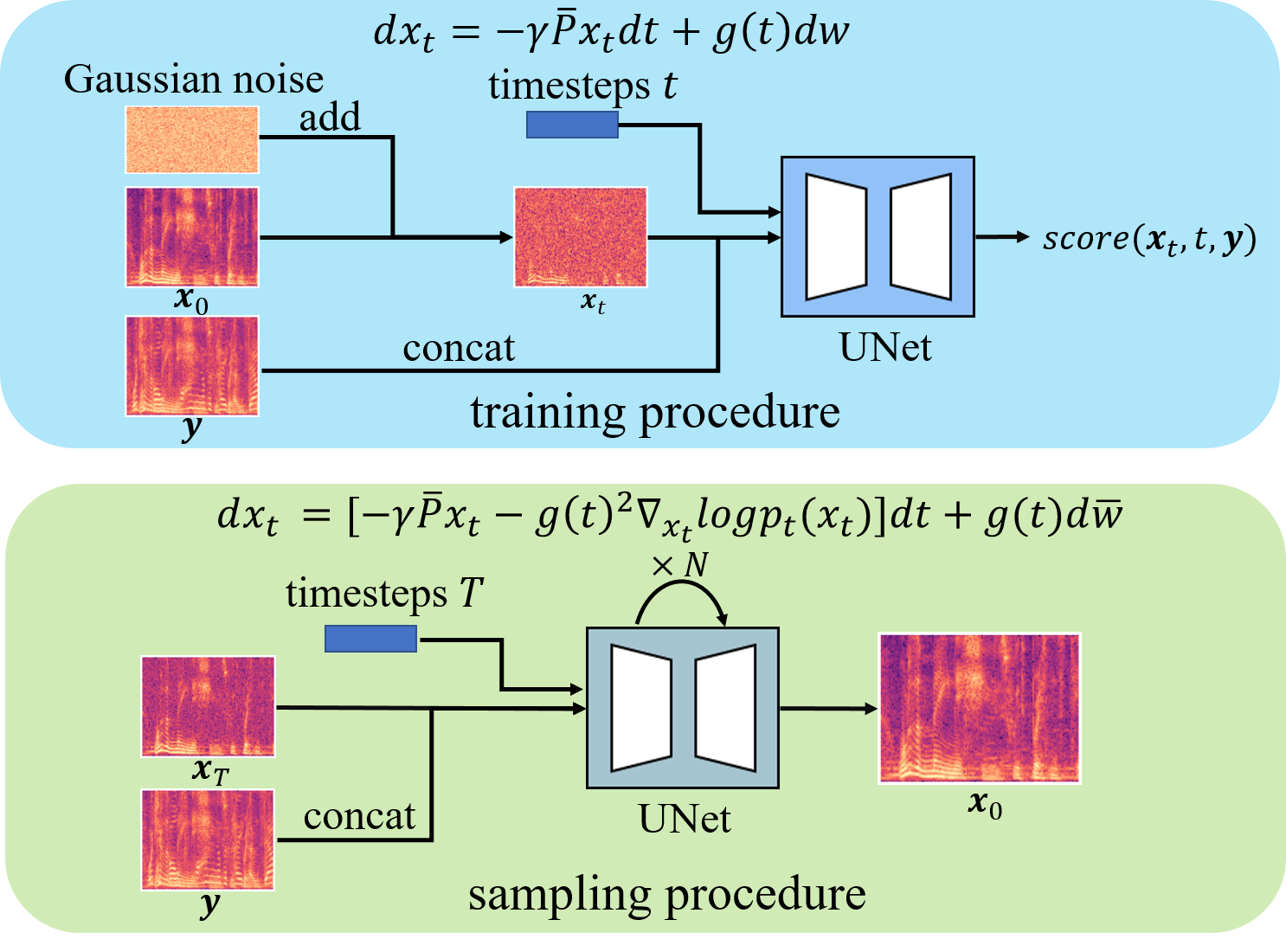}
        \caption{The pipeline of diffusion models for speech separation. In training process, speech sources $\bm{x}_0$ gradually add Gaussian noise and we train the model to approximate the score function. In sampling process, we sample $\bm{x}_{T}$ from the mixture distributions and we use the score-based model to separate mixed speech $\bm{y}$ step by step.}
        \label{fig1}
\end{figure}
Following Scheibler et al. \cite{scheibler2023diffusion}, the forward process can be formalized as the following SDE: 
\begin{equation}
    {d}\bm{x}_{t}=-\gamma\bm{\bar{P}}\bm{x}_{t}dt+g(t)d\bm{w} \textit{,} \quad \bm{x}_0=\bm{s} \textit{,}
\label{eq:SDE_SS}
\end{equation}
\begin{equation}
\bm{P}=K^{-1}\mathbf{E}\mathbf{E}^\top\textit{,}\quad\bm{\bar{P}}=\bm{I}_{K}-\bm{P}\textit{,}
\end{equation}
\begin{equation}
\bm{s}=[{\bm{s}_1}^\top,...,{\bm{s}_K}^\top]^\top\textit{,}\quad\bar{\bm{s}}=K^{-1}[\bm{y}^\top,...,\bm{y}^\top]^\top\textit{,}
\end{equation}
where $\bm{s}$ is the set of the vector of separated sources, $\bar{\bm{s}}$ is their average value, $\bm{y}$ is the speech mixture, $K$ denotes the number of sources, $\mathbf{E}$ is the all-one vector, and $\bm{I}_{K}$ is the $K \times K$ identity matrix. The matrix $\bm{P}$ serves as a projection onto the subspace formed by the average values across sources, while the matrix $\bm{\bar{P}}$ is the orthogonal complement of the space spanned by the matrix $\bm{P}$.

They also formalize the marginal distribution of $\bm{x}_{t}$ as:
\begin{equation}
    \bm{x}_{t}=\bm{\mu}_{t}+\bm{L}_{t}\bm{z}\textit{,} \quad {\bm{z} \sim {\mathcal{N}}(0,\bm{I}_{KM})}\textit{,}
\label{xt}
\end{equation}
\begin{equation}
    \bm{\mu}_{t}=(1-e^{-\gamma t})\bar{\bm{s}}+e^{-\gamma t}\bm{s}\textit{,}
\label{mut}
\end{equation}
\begin{equation}
    \bm{L}_{t}=\sqrt{\lambda_{1}(t)}\bm{P}+\sqrt{\lambda_{2}(t)}\bm{\bar{P}}\textit{,}
\label{Lt}
\end{equation}
\begin{equation}
    \bm{\Sigma}_{t}=\lambda_{1}(t)\bm{P}+\lambda_{2}(t)\bm{\bar{P}}\textit{,}
\end{equation}
\begin{equation}
    \lambda_{k}(t)=\frac{{\sigma_{\min}^2}(\rho^{2t}-e^{-2\xi_{k} t})\log\rho}{\xi_{k}+\log\rho}\textit{,}
\label{lambdat}
\end{equation}
where $\xi_{1}=0$, $\xi_{2}=\gamma$, $\rho=\frac{\sigma_{\max}}{\sigma_{\min}}$, $M$ is the number of samples in the audio signal, $\mu_{t}$ is the mean matrix of $\bm{x}_{t}$ and $\bm{\Sigma}_{t}$ is the covariance matrix of $\bm{x}_{t}$. The SDE in Eq.~\ref{eq:SDE_SS} has the property that the mean of the
marginal $\mu_{t}$ goes from the vector of separated sources as $t=0$ to the mixture vector $\bar{\bm{s}}$ as $t$ grows large. By adjusting the value of $\gamma$, we can reduce the difference between $\mu_{t}$ and $\bar{\bm{s}}$ to an arbitrarily small amount. Fig. \ref{fig1} shows the pipeline of diffusion models for speech separation.

\section{Method}

To enhance the efficiency and quality of diffusion-based speech separation, the SDE in Eq.~\ref{eq:SDE_SS} is adopted with  the diffusion coefficient of the variance exploding SDE~\cite{song2020score}:
\begin{equation}
    g(t)=\sigma_{\min}\Big({\frac{\sigma_{\max}}{\sigma_{\min}}}\Big)^{t} \sqrt{2\log\Big({\frac{\sigma_{\max}}{\sigma_{\min}}}\Big)} \textit{.}
\end{equation}

Previous research has effectively utilized the diffusion process within the mel spectrogram domain incorporating a non-linear transformation~\cite{chen2023sepdiff}. However, due to the inherent nature of the process in Eq.~\ref{eq:SDE_SS} which models the linear mixing of sources, integrating a non-linear transformation directly within this framework is unfeasible. In contrast, our approach implements the diffusion process in the time domain, and our network functions in the domain of the non-linear Short-Time Fourier Transform (STFT).


\subsection{Training Procedure}

Unlike other approaches that train the score network directly, we have adopted specific design choices from the framework of variance exploding SDE~\cite{karras2022elucidating} to design our denoiser function. The denoiser function, similar to the structure in Eq.~\ref{denoiser}, is formulated as:
\begin{equation}
    D_{\theta}(\bm{x}_{t},\sigma(t),\bm{y})={\bm{x}_{t}}+\bm{L}_{t}F_{\theta}({\bm{x}_{t}}, \ln\Big({\frac{1}{2}\sigma(t)}\Big),\bm{y}) \textit{,} 
\end{equation}
where $\bm{x}_{t}$ and $\bm{L}_{t}$ are defined in Eq.~\ref{xt} and Eq.~\ref{Lt}, respectively. We utilize the speech mixture $\bm{y}$ as a conditioning input. To facilitate the model's direct learning of features associated with Gaussian noise, we input the noise level $\sigma(t)$, rather than the time $t$, into the neural network $F_{\theta}$. This noise level is given by:
\begin{equation}
    \sigma(t)=\sqrt{\lambda_{1}(t)}+\sqrt{\lambda_{2}(t)} \textit{,} 
\end{equation}
where ${\lambda_{1}(t)}$ and ${\lambda_{2}(t)}$ are the eigenvalues of the covariance matrix of ${\bm{x}_{t}}$ defined in Eq.~\ref{lambdat}. The training loss is 
\begin{equation}
    \mathcal{L}=\mathbb{E}_{\bm{x}_0, t}{[\Vert {\bm{L}_{t}}^{-1}(D_{\theta}(\bm{x}_{t},\sigma(t),\bm{y})-\bm{\mu}_{t})\Vert^2_2]} \textit{,}
\label{loss}
\end{equation}
where $t\sim\mathcal{U}(t_{0},T)$ and $\bm{x}_0$ are chosen randomly from the dataset. 

If the model is trained exclusively using the previously outlined method, it encounters several significant challenges. Firstly, there is a notable divergence between the expected value $\mathbb{E}[\bar{\bm{x}}_T]=\bar{\bm{s}}$ and $\bm{\mu}_t$,  where $\bar{\bm{x}}_T\sim\mathcal{N}(\bar{\bm{s}},\bm{\Sigma}_{t}\bm{I}_{KM})$. Unlike $\bm{\mu}_t$, which incorporates components representing varying proportions of the sources, $\bar{\bm{s}}$ does not exhibit this variation. Moreover, the network needs to determine the correct sequence in which to present the sources, adding complexity to the task. To address these challenges, we have adapted our training strategy, following Scheibler et al. \cite{scheibler2023diffusion}, to include considerations for model mismatch. For each sample, we set the probability $p_T$, $p_T\in(0,1)$. With the probability $1-p_T$, we apply the standard score matching process, aiming to minimize Eq.~\ref{loss}. With the probability $p_T$, we set $t=T$ and minimize the following training loss:
\begin{equation*}
    \mathcal{L}=\mathbb{E}_{\bm{x}_0}\min_{a\in\mathcal{P}}{[\Vert {\bm{L}_{T}}^{-1}(D_{\theta}(\hat{\bm{x}}_{T},\sigma(T),\bm{y})-\bm{\mu}_{T}(a))\Vert^2_2]} \textit{,} 
\end{equation*}
where $\hat{\bm{x}}_{T}=\bar{\bm{s}}+\bm{L}_{T}$, $\mathcal{P}$ is the set of permutations of the
speech sources
 and ${\bm{\mu}_{T}(a)}$ is $\bm{\mu}_{T}$ computed for permutation $a$.

\subsection{Sampling Procedure}
\begin{algorithm}[t]
\centering
\caption{Our stochastic sampler}
\label{alg:example}
\begin{algorithmic}[1] 
\STATE \textbf{procedure} stochastic sampler($D_{\theta}, t_{i\in\{0,...,N\}}$)
\STATE \textbf{sample} $\bm{x}_{0}\sim{\mathcal{N}}(\bar{\bm{s}},\bm{\Sigma}_{t_0}\bm{I}_{KM})$
\FOR {$i\in\{0,...,N\}$}
    \STATE \textbf{sample} $\bm{n}\sim{\mathcal{N}}(0,\bm{\Sigma}_{t_i}\bm{I}_{KM})$
    \STATE $\bm{\hat{x}}_{i}=D_{\theta}(\bm{x}_{i},\sigma(t_i),\bm{y})+\bm{n}$
    \STATE $\Delta{t}=t_{i+1}-t_{i}$
    \STATE $d_{i}=(-\gamma\bm{\bar{P}}\bm{\hat{x}}_{i}+A\bm{\hat{x}}_{i}-AD_{\theta}(\bm{\hat{x}}_{i},\sigma(t_i),\bm{y}))\Delta{t}$
    \STATE $\bm{{x}}_{i+1}=\bm{\hat{x}}_{i}+d_{i}$
\ENDFOR
\RETURN $\bm{x}_N$
\end{algorithmic}
\end{algorithm}

Inspired by deterministic ordinary differential equation (ODE) integrator \cite{watson2022learning}, we obtain the ODE formulation:
\begin{equation}
    {d}\bm{x}_{t}=(-\gamma\bm{\bar{P}}\bm{x}_{t}+A\bm{x}_{t}-AD_{\theta}(\bm{x}_{t},\sigma(t),\bm{y}))dt \textit{,} 
\end{equation}
\begin{equation}
    A=\frac{{\Dot{\lambda}_{1}}(t)}{{{2\lambda}_{1}}(t)}\bm{P}+\frac{{\Dot{\lambda}_{2}}(t)}{{{2\lambda}_{2}}(t)}\bm{\bar{P}} \textit{,} 
\end{equation}
where the dot denotes a time derivative. The introduction of deterministic sampling by reversing an ODE offers distinct advantages, such as converting training data into their corresponding latent representations. To improve the quality of the separation results, stochasticity, which has been proved useful in\cite{bao2022analytic,jolicoeur2021gotta,song2019generative} is considered.
To solve the Eq.~\ref{eq:SDE_SS}, we propose a stochastic sampler that combines the probability flow ODE with noise injection. A pseudocode is given in Algorithm~\ref{alg:example}. At each step $i$, given the sample $\bm{{x}}_{i}$ at noise level $\sigma(t_i)$, we add noise to the sample before we solve the ODE backward from $t_i$ to $t_{i+1}$ with a single step. The parameters used to evaluate $D_{\theta}$ correspond to the state after noise injection, whereas an Euler–Maruyama method would use $\bm{{x}}_{i}$ instead of $\bm{\hat{x}}_{i}$.

\subsection{Implementation Details}
The noise-conditioned score-matching network (NCSN++) architecture \cite{song2020score} is used as the backbone. Before we input data into the neural network,  we apply the non-linear transform $\operatorname {m}(x)$ \cite{richter2023speech, WelkerRG22} after STFT, and inverse it before inverse STFT. The transform $\operatorname {m}(x)$ is formalized as:
\begin{equation}
    \operatorname {m}(x)=\beta^{-1}|x|^{\alpha}e^{j\angle x}\textit{,}\quad \operatorname {m}^{-1}(x)=\beta|x|^{\frac{1}{\alpha}}e^{j\angle x}\textit{.}
\label{tranformation}
\end{equation}

The SDE parameters are set as follows: $\sigma_{\max}=0.5$, $\sigma_{\min}=0.05$, and $\gamma=2$. For the transformation in Eq.~\ref{tranformation}, $\beta$ and $\alpha$ are set as 0.15 and 0.5, respectively. The probability of selecting a sample is set at $p_T=0.1$. The model is optimized with the Adam optimizer. For the sampling procedure, we set steps $N=29$ which means we use 30 steps for prediction.

\section{Experiments}

\subsection{Datasets}

To evaluate the performance of the proposed method in the task of speech separation, WSJ0-2mix \cite{hershey2016deep}, LRS2-2mix \cite{afouras2018deep}, and VoxCeleb2-2mix \cite{chung2018voxceleb2} are adopted for the speech separation experiment. WSJ0-2mix dataset contains 30 hours of training, 10 hours of validation, and 5 hours of evaluation data. LRS2-2mix dataset contains thousands of BBC video clips, followed by mixing speeches with signal-to-noise ratios between -5 dB and 5 dB. It contains 11 hours of training and 3 hours of validation. VoxCeleb2-2mix dataset is a widely-used dataset for audio tasks comprising more than 1 million utterances extracted from YouTube videos. This dataset consists of 5,994 unique speakers in the training set, and 118 unique speakers in the test set. We only use the audio data in these datasets. The sampling rate is 8 kHz. The length of each speech is 2 seconds. 

\subsection{Results}
\begin{figure}[t]
        \centering
        \includegraphics[width=1\linewidth]{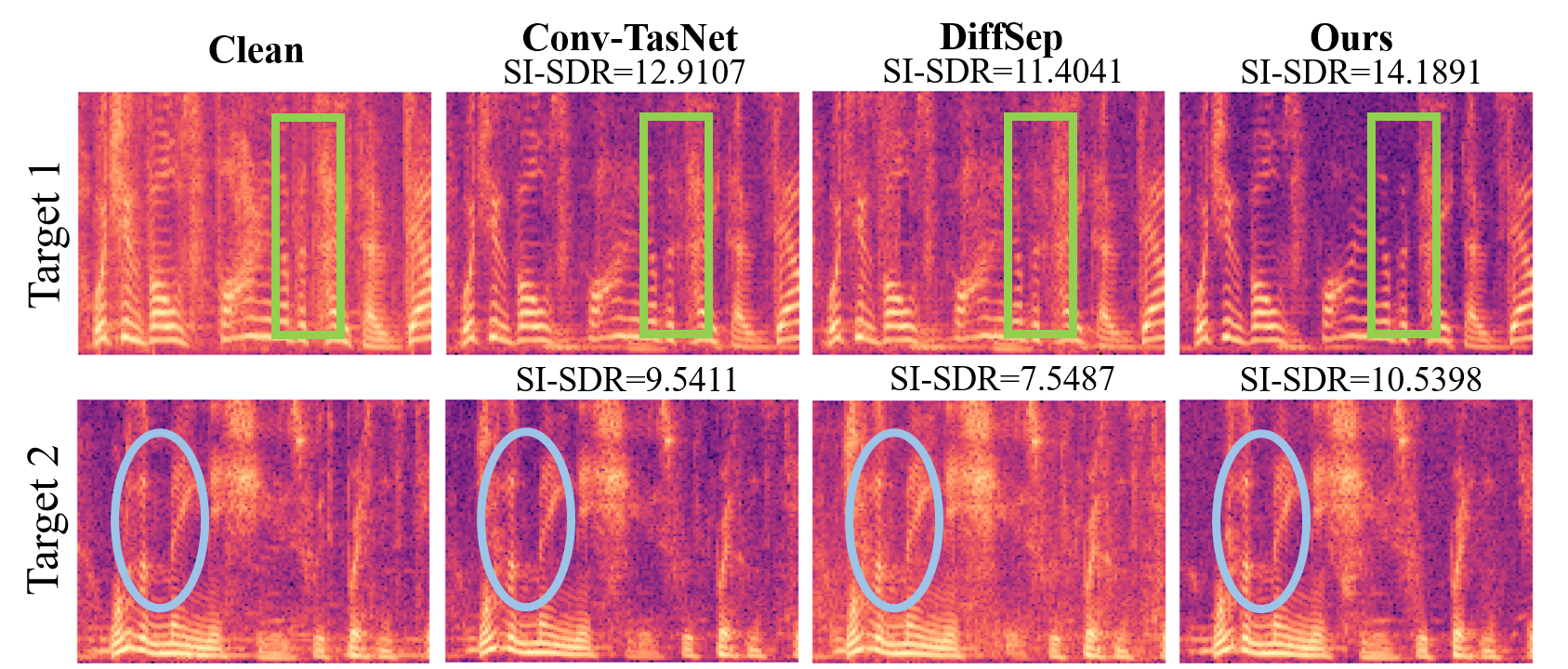}
        \caption{Spectrograms of the clean sources and the separated speech of Conv-TasNet, DiffSep and our method. The sample was chosen randomly from the LRS2-2mix dataset.}
        \label{fig2}
\end{figure}




\begin{table}[!htp]
    \caption{Comparisons of different models in the speech separation task across various datasets.}
    \label{tb:results}
    \centering
    \setlength{\tabcolsep}{2.5mm} 
    \begin{tabular}{lcccc}
    \toprule
    Dataset & Model & SI-SDR & PESQ & ESTOI\\
    \midrule
    \multirow{3}{*}{WSJ0-2mix} & Conv-TasNet\cite{luo2019conv}&15.3&\textbf{3.26}&0.91\\
                                & DiffSep\cite{scheibler2023diffusion}&14.3&3.14&0.90\\
                                & EDSep(ours)&\textbf{15.9}&3.18&\textbf{0.91}\\
    \midrule
    \multirow{3}{*}{LRS2-2mix} & Conv-TasNet\cite{luo2019conv}&9.4&2.43&0.75\\
                               & DiffSep\cite{scheibler2023diffusion}&7.1&2.26&0.67\\
                               & EDSep(ours)&\textbf{9.6}&\textbf{2.51}&\textbf{0.79}\\
    \midrule
    \multirow{3}{*}{VoxCeleb2-2mix} & Conv-TasNet\cite{luo2019conv}&\textbf{7.3}&2.36&\textbf{0.75}\\
                                    & DiffSep\cite{scheibler2023diffusion}&4.9&2.24&0.62\\
                                    & EDSep(ours)&7.1&\textbf{2.45}&0.73\\
    \bottomrule
    \end{tabular}
\end{table}

Evaluation metrics, including SI-SDR \cite{le2019sdr}, perceptual evaluation of speech quality (PESQ) \cite{rix2001perceptual}, and extended short-time objective intelligibility (ESTOI) \cite{jensen2016algorithm}, are used to assess the performance. All of these metrics indicate a better separation performance with a higher value.

In addition to the most recently proposed diffusion-based method DiffSep~\cite{scheibler2023diffusion}, one of the most famous discriminative-based speech separation methods~Conv-TasNet \cite{luo2019conv} is also compared.
The comparative analysis of different models on the speech separation task across various datasets is detailed in Table~\ref{tb:results}. Notably, our proposed model, EDSep, consistently surpasses the diffusion-based method, DiffSep, in all evaluation metrics across all datasets. Specifically, EDSep achieves a 44.9\% higher relative SI-SDR than DiffSep on the VoxCeleb2-2mix dataset, underscoring the robustness of our approach.


In comparison to the discriminative model Conv-TasNet, EDSep exhibits comparable, if not superior, performance. On the LRS2-2mix dataset, EDSep outperforms Conv-TasNet across all metrics. While not universally better on other datasets, EDSep still scores higher in key metrics, demonstrating its effectiveness and competitiveness in speech separation.



To provide a visual demonstration of the different models' separation effects, we compared the spectrograms of audio post-separation, as depicted in Fig.~\ref{fig2}. The spectrogram produced by EDSep offers clearer harmonic information and less noise than the other two models. 
This visual representation illustrates the comparative advantage of the proposed method, highlighting its efficacy and superiority over existing method mentioned in the paper.


\section{Conclusion}

In this work, we introduced EDSep, a speech separation framework leveraging a diffusion model to produce highly natural-sounding separated speech. Our approach significantly enhances the quality of output speech through innovations in the training and sampling processes. Although there is a notable performance disparity with state-of-the-art discriminative methods, the potential for further improvement presents an intriguing challenge. Future work will focus on optimizing training and sampling strategies to close this gap. Additionally, we have yet to explore the theoretical properties of mixture distributions, which may offer further enhancements to our model's performance.


\bibliographystyle{IEEEtran}
\bibliography{IEEEabrv, citations}
\end{document}